# SECURITY ENHANCEMENT WITH OPTIMAL QOS USING EAP-AKA IN HYBRID COUPLED 3G-WLAN CONVERGENCE NETWORK

R. Shankar[1], Timothy Rajkumar K[2] and P. Dananjayan[3]

[1]Department of ECE, Sri Manakula Vinayagar Engineering College, Pondicherry, India
pdananjayan@pec.edu
[2, 3]Department of ECE, Pondicherry Engineering College, Pondicherry, India
pdananjayan@rediffmail.com

## ABSTRACT

*The third generation partnership project (3GPP) has addressed the feasibility of interworking and specified the interworking architecture and security architecture for third generation (3G)-wireless local area network (WLAN), it is developing, system architecture evolution (SAE)/ long term evolution (LTE) architecture, for the next generation mobile communication system. To provide a secure 3G-WLAN interworking in the SAE/LTE architecture, Extensible authentication protocol-authentication and key agreement (EAP-AKA) is used. However, EAP-AKA have several vulnerabilities. Therefore, this paper not only analyses the threats and attacks in 3G-WLAN interworking but also proposes a new authentication and key agreement protocol based on EAP-AKA. The proposed protocol combines elliptic curve Diffie-Hellman (ECDH) with symmetric key cryptosystem to overcome the vulnerabilities. The proposed protocol is used in hybrid coupled 3G-WLAN convergence network to analyse its efficiency in terms of QoS metrics, the results obtained using OPNET 14.5 shows that the proposed protocol outperforms existing interworking protocols both in security and QoS.*

## 1. INTRODUCTION

The third-generation (3G) mobile communication systems provide great coverage, complete subscriber management and nearly universal roaming. Nevertheless, 3G systems [1, 2] are subject to the low data rates. WLAN provides high data rates in short range with less roaming and mobility support. From the users' point of view, the integration of WLAN and 3G systems will provide a convenient and attractive way for the user to access converged network [3, 4]. The users are able to access 3G mobile network while at high speed travel or access WLAN while moving slowly or entering a specific area. However, while integrating WLAN and 3G, there are still some problems that should be considered in terms of security and quality of service. In this paper, Universal Mobile Telecommunications System (UMTS) is focused and the interworking scenarios [5] and architecture of 3G and WLAN are reviewed. The key goal of this integration is to develop convergence network, capable to support ubiquitous data services with very high data rates. The interworking between WLANs and cellular networks is expected to break new ground for sophisticated business models. While 3G systems offer large coverage and a rich network infrastructure [6], WLANs are available in a wide range of devices and are now being viewed as the means of a ubiquitous broadband access platform. 802.11b can provide link speeds of 11 Mb/s and application speeds of 5 Mb/s, and newer standards such as 802.11a/g provide even higher speeds, which is ideal for internet- based data access [7]. An integrated network combines the strengths of each, resulting in a wide area system capable of providing users with ubiquitous data service ranging from low to high speed in strategic locations.

3G-WLAN interworking involves an EAP enabling authentication and session key distribution using AKA mechanism. UMTS-AKA [8, 9] is based on symmetric keys and runs typically on a UMTS Subscriber Identity Module (USIM). EAP-AKA Authentication includes optional user anonymity and re-authentication procedures. EAP-AKA is based on UMTS-AKA. For this reason, EAP-AKA not only have the vulnerabilities of UMTS-AKA but also vulnerabilities in 3G-WLAN interworking. In order to execute the UMTS AKA procedures over EAP, a separate EAP method is defined [10]. EAP-transport layer security (EAP-TLS) is the original, standard wireless LAN EAP authentication protocol. Although it is rarely deployed, it is still considered one of the most secure EAP standards available and is universally supported by all manufacturers of wireless LAN hardware and software. The requirement for a client-side certificate gives EAP-TLS its authentication strength and illustrates the classic convenience vs. security trade-off. A compromised password is not enough to break into EAP-TLS enabled systems because the hacker still needs to have the client-side private key. The highest security available is when client-side keys are housed in smart cards. This is because there is no way to steal a certificate's corresponding private key from a smart card without stealing the card itself. It is significantly more likely that the physical theft of a smart card would be noticed (and the smart card immediately revoked) than a (typical) password theft to be noticed [11]. EAP-tunneled transport layer security (EAP-TTLS) is an EAP protocol that extends TLS. EAP-TTLS offers very good security. The client does not need to be authenticated via a CA-signed PKI certificate to the server, but only the server to the client [12]. This greatly simplifies the setup procedure as a certificate does not need to be installed on every client.

After the server is securely authenticated to the client via its Certification Authority (CA) certificate, the server can then use the established secure connection "tunnel" to authenticate the client. It can use an existing and widely deployed authentication protocol and infrastructure, incorporating legacy password mechanisms and authentication databases, while the secure tunnel provides protection from eavesdropping and man-in-the-middle attack. User's name is never transmitted in unencrypted clear text, thus improving privacy. EAP for GSM subscriber identity is used for authentication and session key distribution using the global system for mobile communications (GSM) subscriber identity module (SIM). GSM is a second generation mobile network standard [13]. The EAP-SIM mechanism specifies enhancements to GSM authentication and key agreement whereby multiple authentication triplets can be combined to create authentication responses and session keys of greater strength than the individual GSM triplets. The mechanism also includes network authentication, user anonymity support, result indications, and a fast re-authentication procedure. EAP-AKA, is an extensible authentication protocol mechanism for authentication and session key distribution in UMTS using USIM [14]. AKA is used in the 3rd generation mobile networks such as UMTS and CDMA2000.

The rest of the paper is organised as follows. Section 2 unveils the authentication and key agreement of EAP-AKA and the proposed protocol. Section 3 deals with coupling scenarios and the proposed coupling architecture. Section 4 presents the simulation model and the simulation results, which depicts the performance analysis of the proposed protocol with the existing protocols. The QoS parameters are considered and the comparison is done with the previous protocol. Section 5 summarises the inferences arrived through the newly proposed protocol. It reveals the merits of the authentication and key agreement of the proposed protocol and its QoS performance.

## 2. AUTHENTICATION AND KEY AGREEMENT

Authentication and key agreement is a security protocol used in 3G networks. AKA is also used for one time password generation mechanism for digest access authentication. AKA is a challenge-response based mechanism that uses symmetric cryptography. AKA provides procedures for mutual authentication of the MS and serving system. The successful execution of AKA results in the establishment of a security association (i.e., set of security data) between the MS and serving system that enables a set of security services to be provided. Major advantages of AKA include larger authentication keys (128-bit), stronger hash function (SHA-1), support for mutual authentication, support for signaling message data integrity, support for signaling information encryption, support for user data encryption, protection from rogue MS. AKA is a mechanism which performs authentication and session key distribution in UMTS networks. AKA is a challenge-response based mechanism that uses symmetric cryptography. AKA is typically run in a UMTS IP Multimedia Services Identity Module (ISIM), which resides on a smart card like device that also provides tamper resistant storage of shared secrets.

## 2.1 Extensible Authentication Protocol

The Extensible authentication protocol (EAP) is best considered as a framework for transporting authentication protocols, rather than as an authentication protocol itself. EAP can be used for authenticating dial-up and VPN connections, and also LAN ports in conjunction with IEEE 802.1X. When the UE attempts to access Non-3GPP such as WLAN, the UMTS-AKA protocol cannot be used. Therefore, EAP-AKA is used to support 3G-WLAN interworking. EAPAKA protocol is based on UMTS-AKA. For this reason, EAP-AKA contains not only vulnerabilities of UMTS-AKA but also vulnerabilities of 3G-WLAN interworking.

## 2.2 Proposed Protocol

In the proposed protocol, a secure channel is established between the AAA server and the HSS. The UE can identify the ID of AAA server and AP in which it is able to access.

### 2.2.1 Elliptic Curve Diffie-Hellman (ECDH)

User A and B publicly agree on an elliptic curve E over a large finite field F and a point P on that curve. The user A and B each selects random number a and b, respectively. Using elliptic curve point-addition, user A and B each publicly compute aP and bP on E. Finally, user A and B each compute abP using private and public values. As a result, solving ECDH is a computationally difficult problem [15]. Generally, most of the previous protocols do not use any kind of public key cryptosystem because UEs have power limitation, low-level computational power, and less storage space. But, technology is significantly improving. For this reason, previous protocols consider use of public key cryptosystems with certificates.

Therefore, the proposed protocol combines ECDH with symmetric key cryptosystem to provide secure communication between 3G and Non-3GPP. ECDH provides the same security properties and uses fewer resources than other public key cryptosystems with certificates. Thus it has less overhead than previous protocols which are based on public key cryptosystems with certificates. In proposed protocol, the UE and the AAA server only stores and manages a, b, aP, and bP .

## 2.3 Comparison

To authenticate WLAN, IEEE 802.1x provides authentication framework based on Extensible Authentication Protocol (EAP). The EAP supports several authentication protocols and each protocol has advantages and disadvantages, respectively. Table 1 shows comparison of proposed protocol with previous protocols. Referring to Table 1, the proposed protocol supports cellular-WLAN interworking and provides strong user identity protection.

Proposed protocol has less overhead than other protocols (EAP-TLS, EAP-TTLS and EAP-SIM) because of using a symmetric key cryptosystem and ECDH. Moreover, proposed protocol prevents man-in-the middle attack and replay attack. In addition, it provides PFS and does not need SQN synchronisation which occurs in EAP-AKA. Therefore, proposed protocol provide more efficient and secure 3G-WLAN interworking than previous protocols.

Table 1. Comparison of proposed protocol with previous protocols.

| | Proposed Protocol | EAP - AKA | EAP – SIM | EAP – TLS | EAP - TTLS |
|---|---|---|---|---|---|
| Type of cryptosystem | Symmetric and ECDH | Symmetric | Symmetric | Public (Certificate) | Public (Certificate) |
| Subscriber Management | Cellular Network Provider | Cellular Network Provider | Cellular Network Provider | WLAN Provider | WLAN Provider |
| Protection of user identity (IMSI) | ✔ | ✗ | ✗ | ✗ | ✔ |
| Cellular - WLAN internetworking | ✔ | ✔ | ✔ | ✗ | ✗ |
| Secure against man-in-the middle attack | ✔ | ✗ | ✗ | ✔ | ✗ |
| Secure against replay attack | ✔ | ✔ | ✔ | ✔ | ✔ |
| Provide PFS | ✔ | ✗ | ✗ | ✗ | ✗ |
| Need for SQN synchronization | – | ✔ | – | – | – |

# 3. COUPLING SCENARIO

The standard developing bodies have attempted to define standards for the interoperation of the two systems and several researchers are seeking the best methods to interwork the two systems. Generally, two solutions have been considered namely, loosely coupled and tightly coupled interworking [18]. The two solutions differ in the level of integration required between the two networks. These solutions are briefly presented in the following section.

## 3.1 Loosely Coupled Interworking

"Loose coupling" is generally defined as the utilisation of WLAN as a packet based access network complementary to UMTS networks. A loosely-coupled architecture allows a WLAN to bypass the 3G core network and interface directly to the core IP network via a WLAN gateway. This approach completely separates the data paths in the WLAN and UMTS networks. Therefore the WLAN data traffic is never injected into the UMTS core network. It allows for independent deployment of WLAN and UMTS networks and roaming agreement can be established between the operators of the two networks. Therefore minimal modifications to UMTS network are required. WLAN Security, mobility and QoS issues are addressed using internet engineering task force (IETF) schemes.

## 3.2 Tightly Coupled Interworking

In tight coupling, the WLAN is directly connected to the UMTS core network. Consequently, all the WLAN traffic is injected into the UMTS core data network. As a result, Support GPRS support node (SGSN) and gateway GPRS support node (GGSN) need to be modified to be able to handle the higher bit rates supported in the WLAN network. The major advantages here are that the IP addresses at the mobile station can be maintained as well as AAA policies and the QoS is guaranteed. This approach is only feasible when both WLAN and UMTS networks belong to the same operator. The complexity and high cost of reconfiguring the 3G core networks and WLAN gateways are major factors that make the tightly coupled approach less competitive. When integrating WLAN and GPRS networks it has to consider that is it suitable to use a standard AAA server that works for both systems, or if separate servers should be used. An AAA server for a GPRS network differs from a WLAN AAA. A combined server that is capable of handling different types of users would be a possible way of integration, thus a shared AAA should have authentication, authorisation, accounting, IP address assignment and mobility management, level of service and billing. Hence the key element in the architecture is the *3G AAA Server*, which is a new functional component incorporated in a 3G PLMN in order to support interworking with WLANs. The 3G AAA Server in the 3G home PLMN terminates all AAA signaling with the WLAN and interfaces with other 3G components, such as the HSS, the home location register (HLR), the charging gateway/charging collection function (CGw/CCF) and the online charging system (OCS). Both the HLR and the HSS are basically subscription databases, used by the 3G AAA Server for acquiring subscription information for particular WLAN MSs.

## 3.3 Methodology Overview

The simulation model is designed in OPNET™ Modeler 14.5. The simulation parameters were selected to accurately model an interworked WLAN-UMTS system. To compare the loose and tight coupling, simulations were done in both architectures with the same simulation factors (number of nodes, traffic).

### 3.3.1 Design Goals

The primary design goal was to interwork WLAN with UMTS so that it can be utilised as an alternate radio access network. A hybrid coupling scheme is proposed to connect UMTS and WLAN [17]. Under the tight coupled system, users of WLAN can also use the services of UMTS with guaranteed QoS and seamless mobility, but it is a problem that the capacity of UMTS core network nodes is not enough to accommodate the bulky data traffic from WLAN, since the core network nodes are designed to handle circuit voice calls or short packets. The proposed coupling scheme differentiates the data paths according to the type of the traffic and can accommodate traffic from WLAN efficiently with guaranteed QoS and seamless mobility. The remaining design goals were intended to focus the design in order to create an open simulation framework with the capabilities to study the issues and trades-offs for interworking WLAN with UMTS.

### 3.3.2 Proposed Network Architecture

A hybrid coupling scheme that differentiates traffic paths according to the type of the traffic is proposed. For the real-time traffic, tightly coupled network architecture is chosen, and for the non-real time and bulky traffic, loosely coupled network architecture is chosen. For example, the traffic generated from SIP would be delivered along the path of APGW-SGSN-GGSN to application server in external PDN, but the traffic like FTP would be forwarded to access routers. The proposed coupling scheme enables to accommodate traffic from WLAN efficiently with guaranteed QoS. In addition, it also guarantees mobility and provides seamless service like the tightly coupled scheme while users are moving or need to change their access technologies. Similar to tight-coupled network, APGW should be added to connect WLAN to UMTS network. The functions of APGW are forward packets to/from AP from/to SGSN or AR, Support Iu_ps-like interface (which is similar to RANAP signaling protocol and Iu data bearer transport), manage radio resources in WLAN and map them onto radio resources in cellular network, set data path to SGSN or AR according to the type of traffic, Differentiate service types: UE sets its service type in the type of service (TOS) or DiffServ code point (DSCP) in the IP header, and APGW checks that and uses to decide path, in which the functions related to differentiating service types and paths are needed additionally for implementing hybrid coupling scheme. In loose-coupled network, real-time traffic could not have appropriate delay and jitter when forwarded through WLAN and internet and because of this the traffic experiences large packet loss or whole packet calls are even blocked. But when connecting WLAN to UMTS based on proposed hybrid coupling scheme, it is possible to support quality of service of real-time traffic and service continuity for WLAN users.

When tight-coupling scheme is applied, packet loss may decrease, but if bulky data traffic like FTP flows into the cellular core network from WLAN, packet loss rate in core network would abruptly increase. But proposed coupling scheme can prevent core network nodes from traffic overflow by means of detouring non-real time traffic of WLAN and not sharing the capacity of core network nodes. When WLAN and UMTS are coupled by proposed scheme, it is possible to support seamless transfer like tight coupling scheme, using inter-RNC handover like procedure. In this case, Iu-bearer can be reused and handover procedure doesn't need set up new bearers, dropping probability and packet loss probability would decrease. And hybrid coupling scheme also supports UMTS services like Location based service and broadcast/multicast service to users in WLAN. To support user's IP mobility for all of two paths of real time and non-real time traffic, it is assumed that the mobile IP and its simultaneous binding option are used. That is, when a UE accesses through WLAN, it belongs to two FAs. Then for real-time traffic of WLAN user, FA in UMTS acts as a FA, and in the other case, FA in WLAN acts as a FA. To use simultaneous binding option, HA should have ability to support this option and a user sets the s-bit in Mobile IP header to 1. Packet transmission procedure is the same as the case without simultaneous binding option. To receive packets, two kinds of operations are possible. First, the user receives the same packets twice from two FAs.  This is the default operation of simultaneous binding, but has a problem to waste network resources. Second operation needs to use TOS or DSCP in IP header. Users set the TOS or DSCP field according to their traffic type, then HA selects one of two FAs based on value of the field and forwards packets to the FA.

### 3.4 Security Issues in UMTS-WLAN

The access security features in UMTS are superset of those provided in GSM. Some new security features are introduced in UMTS to correct the perceived weaknesses of GSM security. It provides mutual authentication between the UMTS subscriber, represented by the USIM, and the UMTS network in the following sense. The serving network confirms the identity of the subscriber while at the same time the subscriber also confirms that he/she is connected to a serving network that is authorized by its home network. UMTS authentication and key agreement (AKA) protocol are described in [10]. In the packet-switch domain of the network, there are three parties communicating in the protocol: the authentication centre (AuC) in the home environment (HE) of the user, the serving GPRS support node (SGSN) and the user represented by one's universal subscriber identity module (USIM). The UMTS AKA protocol uses one secret key, the authentication and key agreement key which is shared between the AuC and the USIM. A primary requirement in 3GPP has been that 3GPPWLAN interworking shall not compromise the UMTS security architecture. Therefore it is required that the authentication and key distribution be based on the UMTS authentication and key agreement (AKA) procedure [16]. The problem of lack of adequate authentication methods for WLAN has led to the development of the IEEE 802.1X standard which is employed in IEEE 802.11i. The 802.1X is a port based network access control standard that uses the extensible authentication protocol (EAP) between the mobile station and access point to perform per session user authentication. EAP messages are encapsulated using EAPoL (EAP over LAN) and sent over wireless links. These EAP messages are then decapsulated at the access point, and then re-encapsulated using RADIUS or DIAMETER protocol for transmission to AAA server. IEEE 802.1x uses standard security protocols, such as RADIUS or DIAMETER, to provide centralised user identification, authentication, authentication, dynamic key management and accounting.

## 4. SIMULATION MODEL

As shown in Figure 1, the simulation model consists of four main parts called *WLAN* that includes access point (WLAN router), AAA server for authentication, WLAN workstation, *UMTS Network which has n*ode B access point, RNC, SGSN, GGSN, HLR for authentication, UMTS workstation, *Internet Provider* (STC) that consist of MMS (multimedia server), FTP server, HTTP server, billing system, router, switch and *Simulation Parameter tools* that has task node to define custom applications like, WLAN authentication which is the different between loose and open coupling where in loose coupling, WLAN authentication will communicate with HLR (five phase's task).

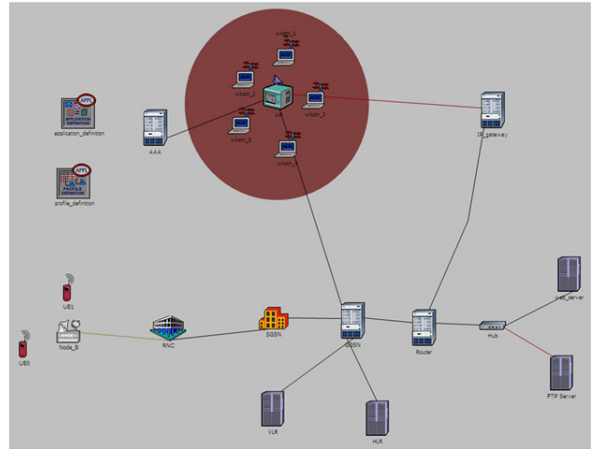

Figure 1. Simulation model

However in open coupling its is normal user ID and password between WLAN user and AAA (three phases task), UMTS authentication which occurs between UE and HLR and it consist of eleven phases,    Multimedia task which communicate with MM server to get MM traffic, application node to define exist applications like HTTP, FTP, Multimedia (MM), billing application as DB traffic, custom task 1: WLAN authentication, custom task 2: UMTS authentication, profile Node to group the applications and assigned them to specific node or user like WLAN authentication, UMTS authentication, WLAN user applications, UMTS user applications, billing task. The model entities have been configured accordingly to generated stated traffic.

### 4.1 Application Response Time

FTP and web application response times were chosen because they belong to two different UMTS QoS profiles. FTP download response time was defined as the time that elapses between a client sending a "get" request and receiving the entire file. It was measured from the time a client application sent a request to the server to the time it received the file. The FTP download response time included the time required to transmit the entire file it was dependent on the file size. Web page response time was defined as the time required retrieving an entire HTML page with all of its inline objects. Similar to the FTP response time, the web page response time was measured from the time a client browser application sent the request to the web server and ends when the client received the entire HTML page.

### 4.2 Traffic Model

The simulation model used the OPNET™ built-in application distribution models. The built-in OPNET application profiles were used to closely simulate traffic generated by a wireless data user. The application profiles used were the FTP, MM, and HTTP profiles. These profiles were combined and parameterized in order to define the wireless client application profile. In addition, since OPNET doesn't support HLR, AAA and Billing system, the traffic for those components was generated via Task tools as described previously. The WLAN and UMTS profiles described a wireless data user's activity over a period of time. The profile consisted of the standard network applications: FTP, MM, and HTTP. Also, authentication and billing profiles were generated to generate AAA, HLR and billing systems. Ten workstations has been created for every access point (WLAN and UMTS). So, there are total of 20 users in the whole network.

### 4.3 Network Design Analysis

As discussed previously, the main difference between open and loose coupling is in the authentication where WLAN have to communication to HLR in loose coupling case. However, for open coupling the authentication is process in every network independently. Therefore, the expectation for is to have more traffic in loose coupling due to AAA-HLR communication during user authentication.

### 4.4 Simulation Results

The simulation is done using OPNET 14.5 Modeler and the results are compared using QoS parameters between the existing EAP-AKA and the proposed, EAP-ECDH protocol.

#### 4.4.1 Wireless LAN Load and Media Access Delay

In Figure 2, the time average in WLAN load is shown in which the wireless LAN load is less in the case of proposed protocol than the existing EAP-AKA. When the load is less the time taken for a process to complete is less in the case of the proposed protocol, the time average in WLAN media access delay is shown where the wireless LAN media access delay is less in the case of proposed protocol than the existing EAP-AKA. When the WLAN media access delay is less, automatically the time taken for access and authentication process to UMTS network or the WLAN is obviously less in the case of the proposed protocol.

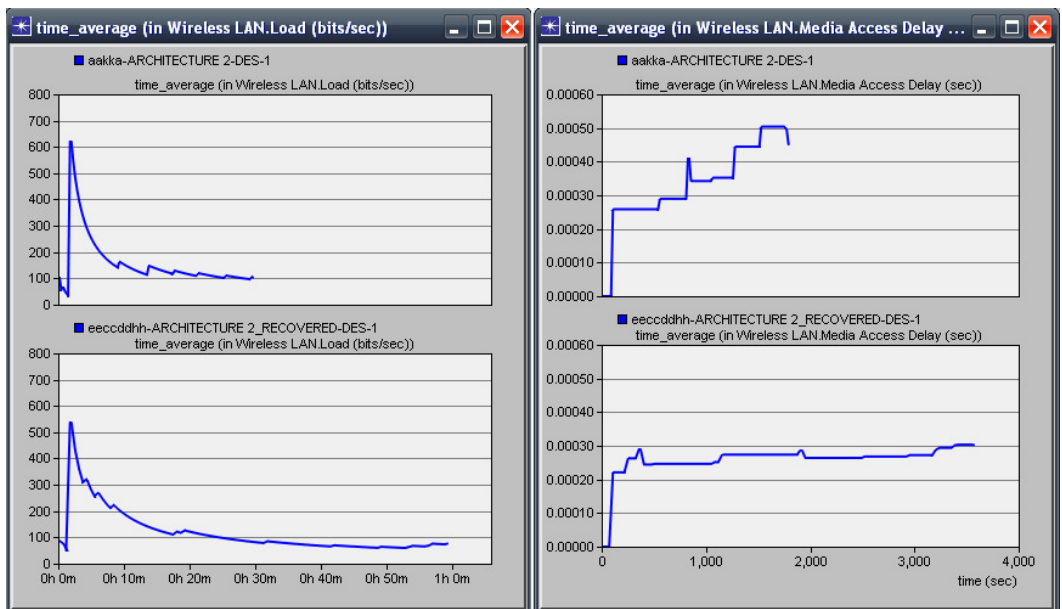

Figure 2. Wireless LAN load and WLAN media access delay

### 4.4.2 Wireless LAN Delay and throughput

In Figure 3, the time average in WLAN delay is shown which shows, the wireless LAN delay is less in the case of proposed protocol than the existing EAP-AKA. When the delay is less automatically the time taken for a process to complete is obviously less in the case of the proposed protocol. In Figure 3, the time average in WLAN throughput [19] is shown that shows, the wireless LAN throughput is more in the case of proposed protocol than the existing EAP-AKA. When the throughput is more the efficiency of the network is quite high in the case of the proposed protocol.

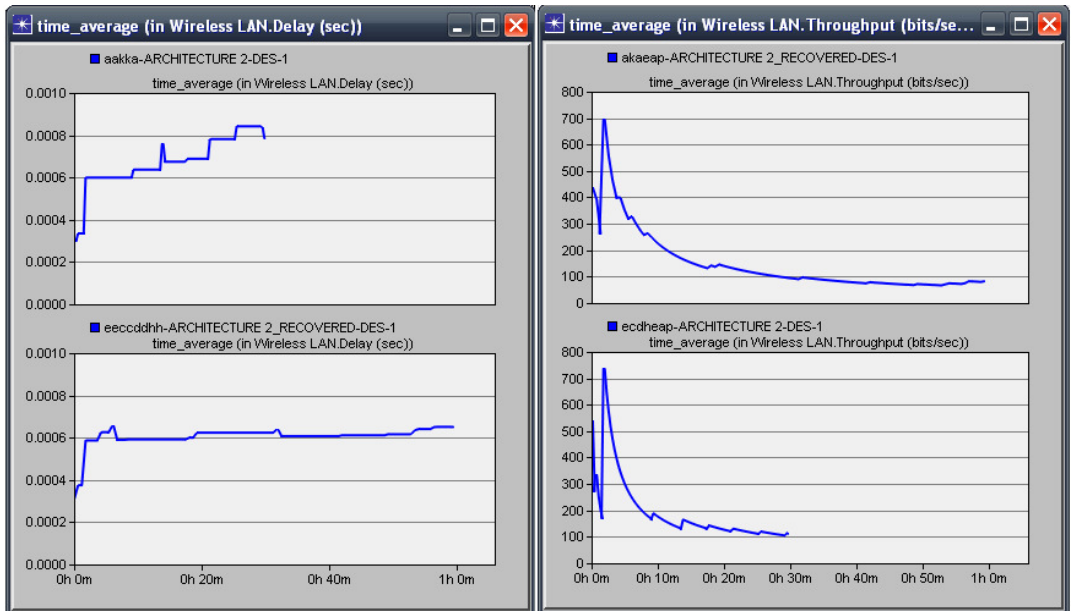

Figure 3. WLAN delay and throughput

### 4.4.3 FTP Traffic and HTTP traffic

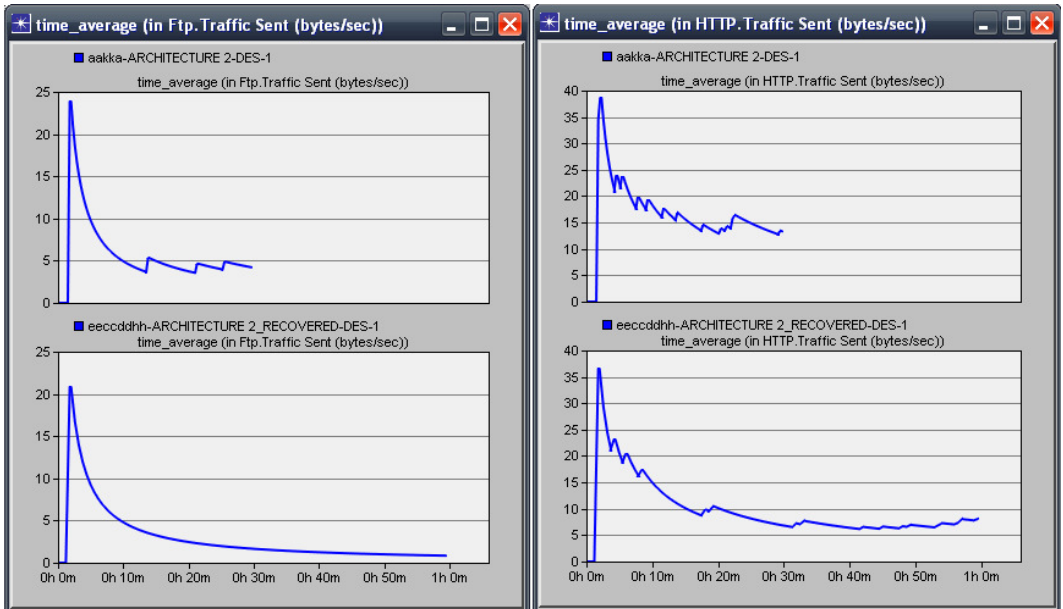

Figure 4. FTP and HTTP traffic

The FTP and HTTP traffic is compared between the existing EAP-AKA and the proposed protocol. In Figure 4, the time average in FTP traffic sent is shown it is found that the FTP traffic is less in the case of proposed protocol than the existing EAP-AKA. When the FTP traffic is less the time taken to send or receive a packet is less in the case of the proposed protocol. In Figure 4, the time average in HTTP traffic sent is shown in which it is observed that the HTTP traffic is less in the case of proposed protocol than the existing EAP-AKA. When the HTTP traffic is less automatically the time taken to send or receive a packet and for a page to respond is obviously less in the case of the proposed protocol.

### 4.4.4 UMTS throughput and load

In Figure 5, the time average in UMTS total received throughput is shown which shows that the UMTS throughput is more in the case of proposed protocol than the existing EAP-AKA. When the throughput is more the efficiency of the network is high in the case of the proposed protocol. Finally the UMTS load of proposed protocol is compared with the existing EAP-AKA In Figure 5, the time average in UMTS total transmit load is shown, here the UMTS load is less in the case of proposed protocol than the existing EAP-AKA. When the load is less the time taken for a process to complete is very less in the case of the proposed protocol.

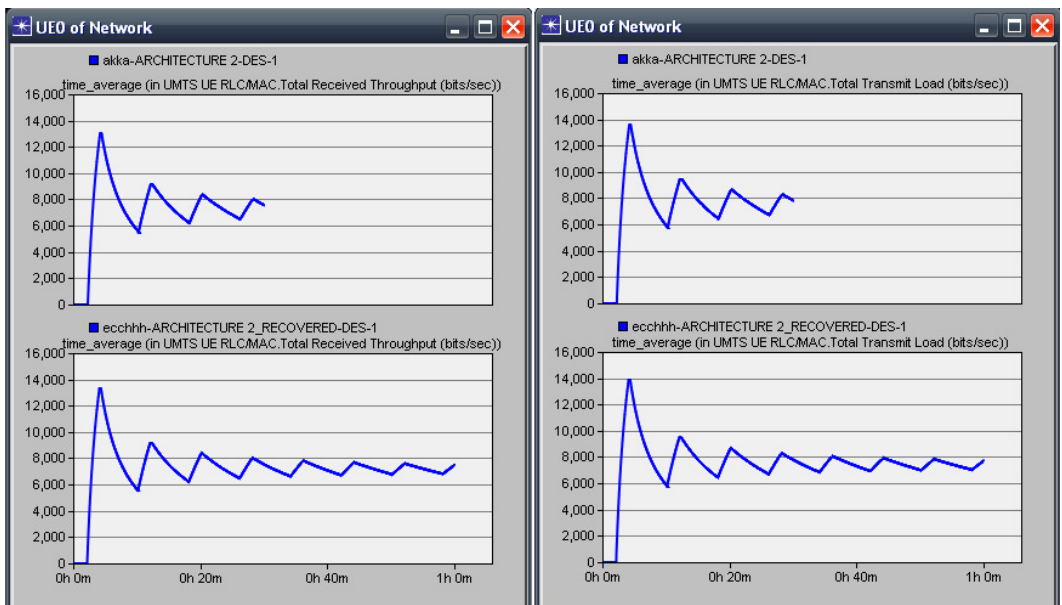

Figure 5. UMTS throughput and load

## 5. CONCLUSION

In this paper, the threats and attacks in 3GWLAN interworking is analysed and a new authentication and key agreement protocol based on EAP-AKA is proposed. It also assures that the proposed protocol provides good quality of service and seamless mobility. Most of the previous protocol does not use any kind of public key cryptosystem because UEs have power limitation, low-level computational power and less storage space. However, technology is significantly improving. For this reason, previous protocols consider the use of public key cryptosystems with certificates. Therefore, the proposed protocol provides secure communication between 3G and Non 3GPP. ECDH provides the same security properties and uses fewer resources than other public key cryptosystems with certificates. The proposed protocol combines ECDH with symmetric key cryptosystem to overcome several vulnerabilities of EAP-AKA such as disclosure of user identity, man-in-middle attack, SQN synchronisation and additional bandwidth consumption. Moreover, proposed protocol provides PFS to guarantee stronger security, mutual authentication between the UE and the AAA server and between the UE and the HSS, and resistance to replay attack. Compared with previous protocols which use public key cryptosystem with certificates, the proposed protocol can reduce computational overhead.

In fact, there is more traffic that needs to go between WLAN and UMTS for authentication purpose. However, the proposed coupling scheme strengthens the concept of integrating WLAN with UMTS by sharing authentication source. Implementing the proposed coupling method will reduce the maintenance cost and will maintain the authentication in one entity instead of two with the cost of extra traffic. Therefore, there is a trade-off between integration advantage and extra network traffic. The proposed protocol also provides good quality of service when compared to the previous existing protocol. Therefore in WLAN, the load, media access delay and delay are less, in FTP and HTTP, the traffic is less, in UMTS and the load is less in the proposed protocol than that of the existing protocol. It is also proved that the WLAN throughput and UMTS throughput is quite high in the case of the proposed protocol thereby providing optimum QoS with better security.

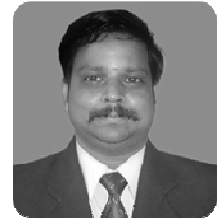

**R. Shankar** received Bachelor of Engineering in 2001 and Master of Technology in 2006 in Electronics and Communication Engineering (ECE) from Bharathidasan University, Trichy and Pondicherry Engineering College, Pondicherry, India respectively. He is pursuing his Ph.D. programme in the Department of ECE, Pondicherry University. He is currently working as Assistant Professor in the Department of ECE at Sri Manakula Vinayagar Engineering College Pondicherry, India. His research interests include wireless communication, computer networks and convergence network.

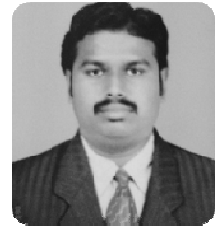

**K. Timothy Rajkumar** received Bachelor of Technology in Information Technology from Pondicherry University. He is pursuing his M.Tech programme in Wireless Communications in the department of Electronics and Communication Engineering at Pondicherry Engineering College, Pondicherry. His area of interest includes 3G-WLAN interworking, Security in Wireless networks, Convergence networks and Computer networks.

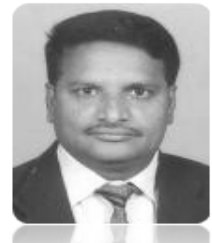

**P. Dananjayan** received Bachelor of Science from University of Madras in 1979, Bachelor of Technology in 1982 and Master of Engineering in 1984 from the Madras Institute of Technology, Chennai and Ph.D. degree from Anna University, Chennai in 1998. He is working as a Professor, Department of ECE, Pondicherry Engineering College, India. He also has been as visiting Professor to Asian Institute of Technology, Thailand. He has more than 60 publications in National and International Journals. He has presented more than 130 papers in National and International conferences. He has guided 9 Ph.D candidates and is currently guiding 6 Ph.D students. His areas of interest include spread spectrum techniques and wireless communication, wireless adhoc and sensor networks.